\documentclass[journal]{IEEEtran}
\usepackage{graphicx}
\usepackage{mathptmx}

\usepackage{color}
\begin{document}

\title{Solar Flare Measurements with STIX and MiSolFA}

\author{{\Large Diego~Casadei}%,~\IEEEmembership{Member,~IEEE}
  \\{FHNW, School of Engineering, Bahnhofstrasse 6, CH-5210
    Windisch, Switzerland.}%
  \thanks{Invited talk, N30-8, Astrophysics and Space Instrumentation
    session, 2014 Nuclear Science Symposium and Medical Imaging
    Conference, 11 Nov 2014.}%
}

\maketitle

\begin{abstract}
  Solar flares are the most powerful events in the solar system and
  the brightest sources of X-rays, often associated with emission of
  particles reaching the Earth and causing geomagnetic storms, giving
  problems to communication, airplanes and even black-outs. X-rays
  emitted by accelerated electrons are the most direct probe of solar
  flare phenomena. The Micro Solar-Flare Apparatus (MiSolFA) is a
  proposed compact X-ray detector which will address the two biggest
  issues in solar flare modeling. Dynamic range limitations prevent
  simultaneous spectroscopy with a single instrument of all X-ray
  emitting regions of a flare. In addition, most X-ray observations so
  far are inconsistent with the high anisotropy predicted by the
  models usually adopted for solar flares. Operated at the same time
  as the STIX instrument of the ESA Solar Orbiter mission, at the next
  solar maximum (2020), they will have the unique opportunity to look
  at the same flare from two different directions: Solar Orbiter gets
  very close to the Sun with significant orbital inclination; MiSolFA
  is in a near-Earth orbit. To solve the cross-calibration problems
  affecting all previous attempts to combine data from different
  satellites, MiSolFA will adopt the same photon detectors as STIX,
  precisely quantifying the anisotropy of the X-ray emission for the
  first time. By selecting flares whose footpoints (the brightest
  X-ray sources, at the chromosphere) are occulted by the solar limb
  for one of the two detectors, the other will be able to study the
  much fainter coronal emission, obtaining for the first time
  simultaneous observations of all interesting regions. MiSolFA shall
  operate on board of a very small satellite, with several launch
  opportunities, and will rely on moir\'e imaging techniques.
\end{abstract}

%\begin{IEEEkeywords}
%IEEEtran, journal, \LaTeX, paper, template.
%\end{IEEEkeywords}

 \section{Introduction}

 \IEEEPARstart{S}{olar} flares are the most powerful events in the
 solar system, releasing $10^{25}$--$10^{26}$~J in few minutes
 \cite{benz2008}.  A large fraction of this energy, initially stored
 in magnetic fields in the solar corona, goes into the acceleration of
 electrons and ions.  Coronal mass ejections (CMEs), often associated
 with solar flares, may reach the Earth, produce geomagnetic storms,
 and cause severe problems to communication, airplanes and, in special
 cases, even large-scale black-outs.

\begin{figure*}
   \centering
   \includegraphics[height=6.5cm]{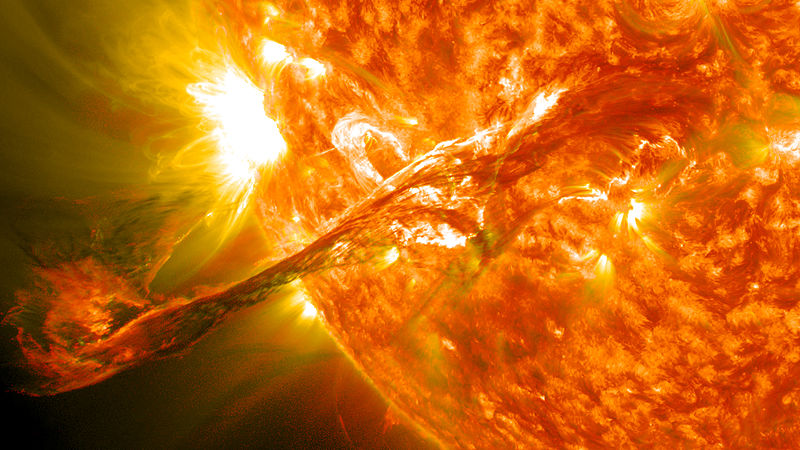}
   \hspace{7mm}
   \includegraphics[height=6.5cm]{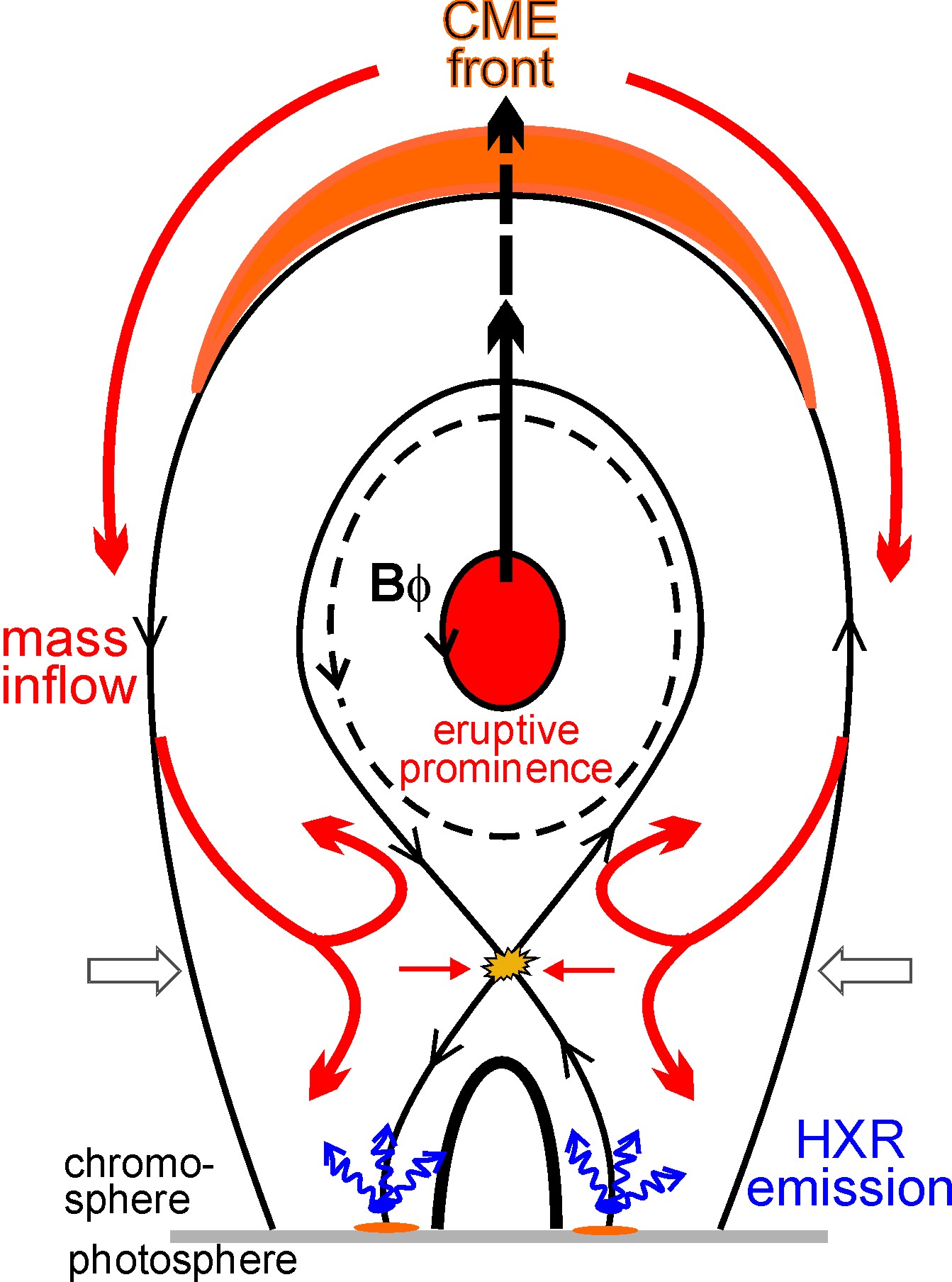}
   \caption{Solar flare with CME on 2012-08-31 15:22UTC (left) and
     illustration of a flare and the associated CME (right).}
   \label{fig-cme}
 \end{figure*}

 The details of the mechanisms transforming magnetic into kinetic
 energy are not yet understood.  The hard X-rays (HXR) emitted by
 electrons accelerated during a flare are the most direct probe of the
 physical processes at the Sun.  The brightest X-ray sites are the
 footpoints of magnetic loops which connect the acceleration regions
 (higher in the corona; Fig.~\ref{fig-cme}) with the much denser
 chromosphere, where the electrons interacting with the ambient ions
 emit bremsstrahlung photons, loosing most of their energy.  From the
 measured photon spectrum, one can infer the distribution of the
 parent electrons, typically a superposition of a thermal component
 with a non-thermal distribution extending to higher energies. Because
 the Earth atmosphere absorbs X-rays, the measurements need to be
 carried out in space.

 % \begin{figure}
 %   \centering
 %   \includegraphics[width=0.6\columnwidth]{flare_cartoon.jpg}
 %   \caption{Illustration of a flare and the associated CME.}
 %   \label{fig-cme}
 % \end{figure}

\begin{figure}%[b]
   \centering
   \includegraphics[width=0.6\columnwidth]{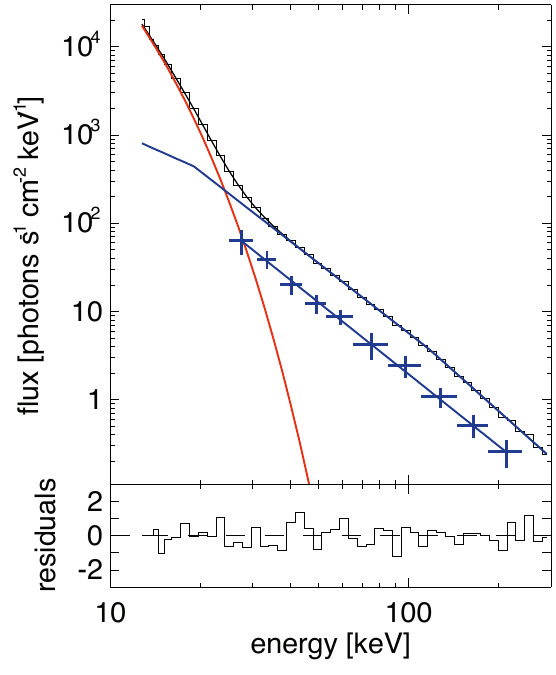}
   \caption{The RHESSI measurement of the overall spectrum from the
     bright (X6.5) flare 2006-12-06 18:47UTC shows a thermal (red) and
     non-thermal (blue; broken power-law) component.  The points with
     error bars are the spectrum from the main footpoint, with a
     single power-law fit \cite{krucker2011}.}
   \label{fig-spectrum}
 \end{figure}

 The most sensitive solar HXR observatory to date is the Reuven Ramaty
 High Energy Solar Spectroscopic Imager (RHESSI \cite{lin2002}), using
 9 rotation modulation collimators and bulk Ge detectors to perform
 imaging spectroscopy from 3 keV to 17 MeV with angular resolution of
 few arcseconds \cite{hurford2002}.  The X-ray energy spectrum from a
 solar flare typically shows a thermal component and a non-thermal
 tail extending possibly to very high energies
 (Fig.~\ref{fig-spectrum}).  The flux of photons quickly falls down
 with increasing energy, typically as a power-law with spectral index
 in the range from 2 to 5 units for the footpoints, and from 3 to 8
 for the coronal source \cite{KL2008}.

 Thanks to RHESSI measurements, it was discovered that 10--50\% of the
 energy release goes into accelerating electrons \cite{emslie2012},
 and that a coronal non-thermal component is present in essentially
 all flares \cite{KL2002}.  Double coronal sources suggest
 acceleration in the corona \cite{SH2003}, and rare observations of
 coronal acceleration sites suggest energization of practically the
 entire pre-existing electron population \cite{krucker2010}.  Hence
 understanding the dynamics of a solar flare requires measurements of
 the photon spectra emitted by the top-of-the-loop region together
 with the footpoints.

 During the next solar maximum (beyond the end of the RHESSI mission),
 the Spectrometer/Telescope for Imaging X-rays (STIX) instrument
 \cite{krucker2013} of the ESA M-class Solar Orbiter mission
 \cite{mueller2013} will allow the scientific community to continue
 studying the X-ray emission from solar flares with similar imaging
 spectroscopy performance as RHESSI, although in a smaller energy
 range (up to 150 keV).  Onboard the Solar Orbiter, additional
 instruments will perform in-situ measurements of solar wind,
 energetic particles, magnetic field and radio and plasma waves, while
 others perform remote measurements in different wavelengths, from
 optical to EUV.  Its orbit will pass closer to the Sun than Mercury
 (with 0.28 AU closest perihelion), allowing STIX to observe it from a
 wide range in latitude (with 25 deg orbit inclination).  STIX exploits
 the moir\'e patterns produced by 30 pairs of grids with different
 periods and relative angle, in order to provide good image quality
 without a rotating satellite.

 Despite from the excellent performance of RHESSI and STIX, solar
 flares are challenging objects of study for any single HXR
 instrument.  For this reason, we have proposed to build a small
 detector called MiSolFA (the Micro Solar-Flare Apparatus), orbiting
 the Earth during the next solar maximum, which together with STIX
 will obtain qualitatively new information about solar flares.
 Together, STIX and MiSolFA will be able to address the two main open
 issues about solar flares, performing a simultaneous measurement of
 coronal sources and footpoints, and evaluating the X-ray emission as
 a function of the direction and energy.

 \section{Open issues}

 Because the bremsstrahlung intensity is proportional to the ambient
 plasma density, the coronal sources (where the acceleration is
 thought to occur) are 1--2 orders of magnitude fainter than the
 bright chromospheric footpoints of flare loops (where the electrons
 loose most of their energy).  The indirect imaging technique based on
 the shadows projected by absorbing gratings on the detector, adopted
 both by RHESSI and STIX (albeit with different methods), has
 limitations in dynamic range: a coronal source is practically
 invisible when a footpoint is inside the field of view of the
 instrument.  Although grazing incidence optics do not have these
 dynamic range limitations, they can not cover the highest energy
 range and require very long satellites (the focal distance being of
 order 10~m), resulting in much higher mission costs.  At present, no
 dedicated mission is foreseen with these optics for solar flare
 studies.

 So far, the best results have been obtained by studying the solar
 flares occurring just behind the solar limb, with occulted footpoints
 but unobscured coronal sources (a few per month during a solar
 maximum).  Statistical studes of partially occulted flares have shown
 that nearly all contain non-thermal coronal sources \cite{KL2008},
 but so far the simultaneous observation of footpoints and coronal
 sources was not possible for any solar flare with the same technique
 (there were observations from different satellites \cite{McTP1990,
   kane1998}, but systematic effects made it impossible to properly
 cross-correlate them).

 Another problem in understanding solar flares is that most
 acceleration models produce strongly beamed non-thermal electron
 fluxes, resulting in highly non-uniform patterns of photon emission.
 However, there is still no conclusive proof that the emitted photon
 spectrum is not uniform.  So far, HXR directivity has been studied
 with the following techniques:
 \begin{itemize}
 \item statistical studies of center-to-limb variations in flux or
   spectral index \cite{ohki1969, kasparova2007} 

 \item analysis of contributions from reflected radiation in a single
   flare \cite{DK2013}

 \item simultaneous observations with different satellites
   \cite{McTP1990, kane1998} 

 \item polarization measurements with a single satellite \cite{mcconnell2004}.
 \end{itemize}

 Many, although not all, these studies have found results consistent
 with an isotropic electron distribution, inconsistent with the strong
 beaming predicted in the thick-target model used to model the
 footpoint emission.  Thus either we lack good observations or we fail
 to understand the fundamental theory.  The only way of solving this
 issue is to directly measure the HXR directivity with
 cross-calibrated detectors observing the same flare from different
 directions.

 Thus, the best option is to have the same kind of instruments
 observing a solar flare from different points of view.  This way, the
 energy response will be the same, which is a crucial requirement.
 Indeed, the photon spectrum is so steep that a small change in the
 measured energy makes a big impact on the overall slope.  On the
 other hand, cross-calibrated instruments are able to measure the
 ratio between the intensities along different lines in each energy
 bin, providing for the first time strong constraints on the emission
 models. 
 For this reason, MiSolFA adopts the same photon detectors as STIX:
 the instrument consists of an array of Caliste SO units equipped with
 CdTe crystals \cite{calisteSO}.  For the solar flares whose
 footpoints are occulted by the solar limb for MiSolFA but visible by
 STIX (or vice versa) it will be possible to measure the energy
 spectrum of the photons coming from both the corona and the
 chromosphere.  In addition, the emissivity of a source visible by
 both instruments will be estimated along two different directions as
 a function of the photon energy.

 \section{The MiSolFA detector}

 The main scientific goals of the MiSolFA project, ranked by
 importance, are:
 \begin{enumerate}
 \item providing together with STIX simultaneous observations of HXR
   from coronal and footpoint sources with cross-calibrated detectors,
   for the first time;

 \item unambiguously measure for the first time the directivity of HXR
   emission together with STIX;

 \item extend the sample of solar flare data with stand-alone
   observations, when STIX is outside its science windows or it is on
   the other side of the Sun.
 \end{enumerate}
 For the subset of CMEs reaching the Earth and detected by MiSolFA, it
 will also possible to perform a detailed study of the magnetic
 connectivity from the Sun to the Earth (one of the 4 top-level
 scientific goals of the Solar Orbiter \cite{mueller2013}).

 There are two possible configurations, one which shall fully achieve
 all science goals, and a fallback design which is technically less
 challenging while preserving most of the science.  In the default
 configuration, X-ray imaging will be indirectly performed along the
 same lines as STIX, although with less details.  This is achieved by
 placing two opaque grids in front of each Caliste unit, producing
 moir\'e patterns whenever a source not much larger than their angular
 scale (half of the ratio between the grid distance and the grating
 period) is visible.  The fallback configuration omits the grids and,
 possibly, reduces the number of Caliste units.  This implies a
 reduction in the number of useful flares for the first two science
 goals, but does not preclude their successful achievement, as even a
 single well-observed event would be sufficient to perform a
 significant step forward in our understanding of solar flares.  The
 biggest impact is on the third goal, as imaging adds qualitatively
 important information to flares observed by MiSolFA alone (for
 example, it allows for a direct estimation of the source volume).
 Still, when no other instrument can detect the same event, even the
 overall spectrum alone brings important information.  In the
 following, only the full configuration is illustrated.

 Imaging is performed indirectly: in front of each Caliste unit, two
 opaque grids will project shadows forming a moir\'e pattern, whose
 period matches the detector size and runs along its pixels.  The
 amplitude of each moir\'e pattern depends on the source size, while the
 phase depends on the source direction. Hence each pair of grids forms
 a subcollimator providing information along one direction (orthogonal
 to the slits) with angular resolution given by (half of) the ratio
 between the grating period and the separation between front and rear
 grids.

 While for RHESSI and STIX a driver in the design was the ability to
 span a wide range of angular scales---before RHESSI they were
 poorly known; STIX will operate in a wide range of distances from the
 Sun---, this is not an important requirement for MiSolFA.  Hence it
 was preferred to give priority to the image resolution at the expense
 of the range of angular scales.  The first requirement is to separate
 the footpoints, which translates into a smallest angular scale of
 about 10 arcseconds.  On the opposite side, one wishes to be able to
 recognize the overall topology of the HXR emission, which means that
 the detector should cover an angular scale of at least 60 arcseconds.
 The best image resolution is then achieved by sampling over integer
 multiples of the baseline angular frequency, as this produces the
 first terms of a Fourier series.  For this reason, the angular scales
 are chosen equal to $60''/n$ with $n=1,2,\ldots,6$. 

\begin{figure}%[b]
   \centering
   \includegraphics[width=0.9\columnwidth]{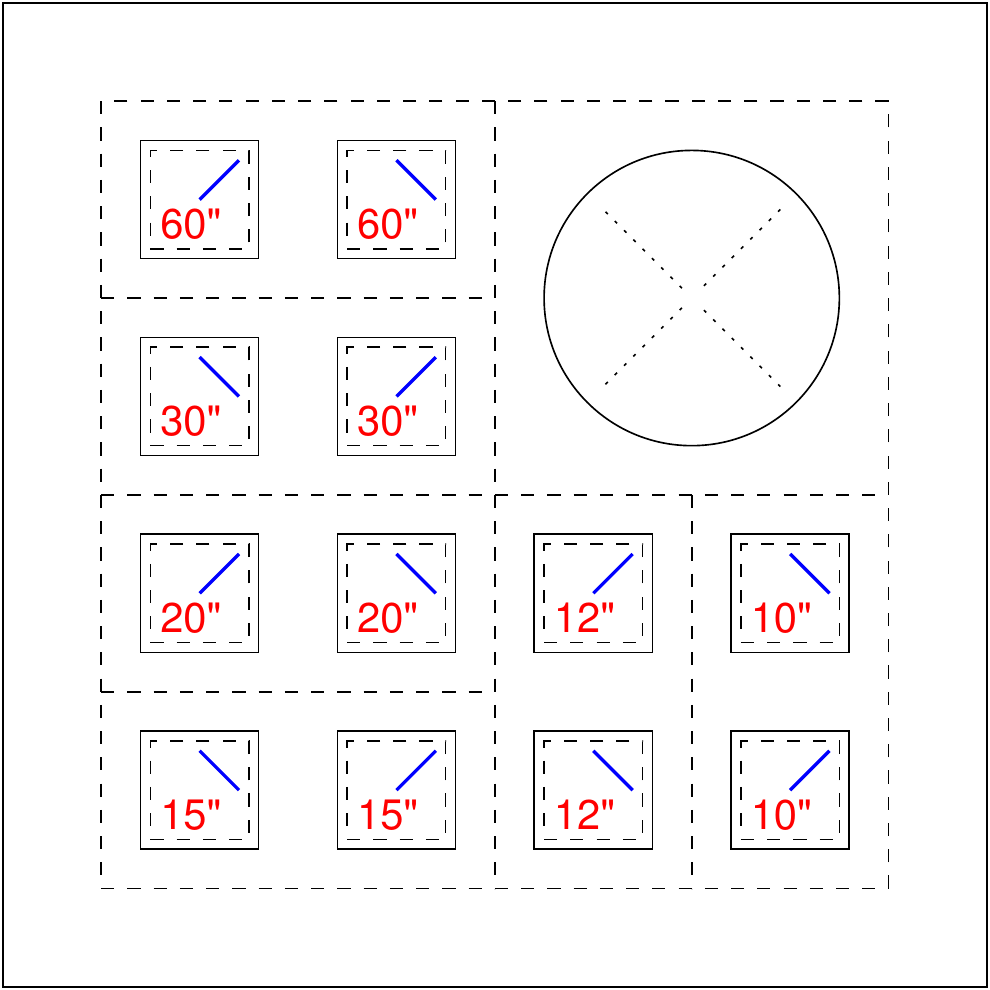}
   \caption{Layout of MiSolFA detectors, viewed from the Sun.}
   \label{fig-layout}
 \end{figure}

 Next, as it is easier to keep the pointing to the required stability
 (within 1 degree) when the satellite is not rotating, one should take
 care of sampling two orthogonal directions at each angular scale.
 This means that the total number of subcollimators is 12, as shown in
 Fig.~\ref{fig-layout}.  As the slit-to-period ratio is 0.5, the
 overall transmission is 0.25 at low energy (increasing toward 100\%
 at high energies, when the grids are no more opaque), giving 3~cm$^2$
 effective area for the thermal component (representing the bulk of
 the flare emission).

 \begin{figure*}
   \centering
   \includegraphics[width=0.8\columnwidth]{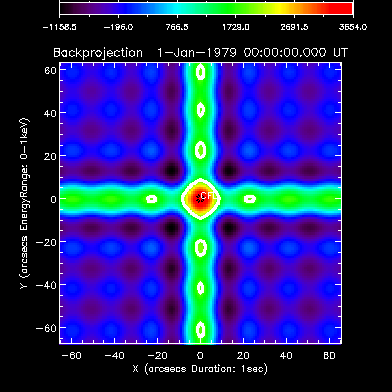}
   \hspace{0.2\columnwidth}
   \includegraphics[width=0.8\columnwidth]{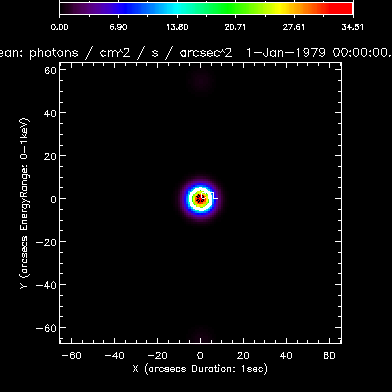}
   \caption{Point spread function of MiSolFA, with the backprojection
     (left) and clean (right) algorithms.}
   \label{fig-psf}
 \end{figure*}

 Figure~\ref{fig-psf} shows the point spread function (PSF) of
 MiSolFA, which corresponds to the reconstructed image of a simulated
 point source at the center of the field of view, using two
 algorithms: backprojection and clean \cite{hurford2002}.  The
 backprojected image is obtained by summing, for each photon, its
 arrival probability distribution.  The photon is detected by one of
 the detectors, which means that it had to cross the two grids in
 front of it.  At low energy, when the grids can be considered opaque,
 this probability distribution is a collection of excluded areas
 mimicing the ``shadow'' of the two grids, intermixed with areas with
 non-null and uniform probability (neglecting diffraction effects).
 At energies for which the transparency of the grids is sizable, the
 ``shadow'' corresponds to areas with small but not null probability.
 By summing over the probability distributions of the photons detected
 by all subcollimators, one gets the position of the source, with a
 typical spread which has a cross shape.  These beams may be removed
 using the clean algorithm, which starts from the backprojected image
 and refines it iteratively, by removing the brightest spot and
 recomputing the image at each step.  At the end of the loop, all
 spots are then summed together, producing an image which is free from
 the ``noise'' visible with the backprojection.

 MiSolFA images are actually periodic structures: they repeat every
 120 arcsecods, as the largest angular scale is $60''$.  Hence MiSolFA
 has to rely on some other observation to provide the precise location
 of the flare in the field of view (which contains the full solar
 disc).  This compromise is not a real problem, as there are may ways
 of obtaining this position using data from other observatories.  To
 uniquely identify the position of the flare would require adding
 angular scales up to about thousand arcseconds (similar to STIX),
 which can not be achieved in a very small volume.  Thus, priority was
 given to the image resolution at the price of some unimportant
 ambiguity in the exact location.

 The precise position of the Sun inside the field of view of the
 instrument is provided by an Aspect System (visible at the top-righ
 corner in Fig.~\ref{fig-layout}) derived from STIX: a lens focusing
 the image of the Sun onto a plate with apertures of different area,
 viewed by photodiodes.  Each of the 4 radial sequences of apertures
 is viewed by a single photodiode.  Their area increases in
 log-uniform way when moving out from the center of the field of view.
 When the Sun image moves on the focal plane, it illuminates a
 variable number of apertures, causing the total intensity measured by
 each diode to change in steps which also follow a log-uniform scale.
 This keeps resolution in pace with the dynamic range, and allows to
 follow the jitter of the optical axis as a function of time within
 few arcseconds.

 The choice of similar subsystems for the photon detection and the
 determination of the Sun position, in addition to guarantee
 cross-calibration, is also a way of insure the space qualification of
 all critical components, of exploiting the close collaboration of
 several STIX members, and of reducing the development time.  The
 synergies between the two projects are a very valuable resource.
 Indeed, the MiSolFA collaboration presently includes the STIX teams
 in Switzerland (FHNW Windisch and PSI), France (CEA Saclay) and Italy
 (University of Genova).

 Assuming that there is space enough for a 20 cm separation between
 the grids, the period of all grids is fixed by the series of angular
 scales mentioned before.  The smallest period in MiSolFA is about
 20~microns (for $10''$), beyond the current state of the art of grid
 fabrication for space applications (based on chemical etching of
 tungsten guided by lithography).  Hence it is planned to adopt a
 different fabrication technique, which shall be finally able to
 create gold structures as small as 1--2 microns (20--40 times smaller
 than with chemical etching).  The PSI group and microworks in
 Karlsruhe, Germany, are currently trying to create gratings for
 phase-contrast radiography \cite{david2007} and are willing to
 produce the grids for MiSolFA.

 \section{The satellite}

\begin{figure}
   \centering
   \includegraphics[width=0.6\columnwidth]{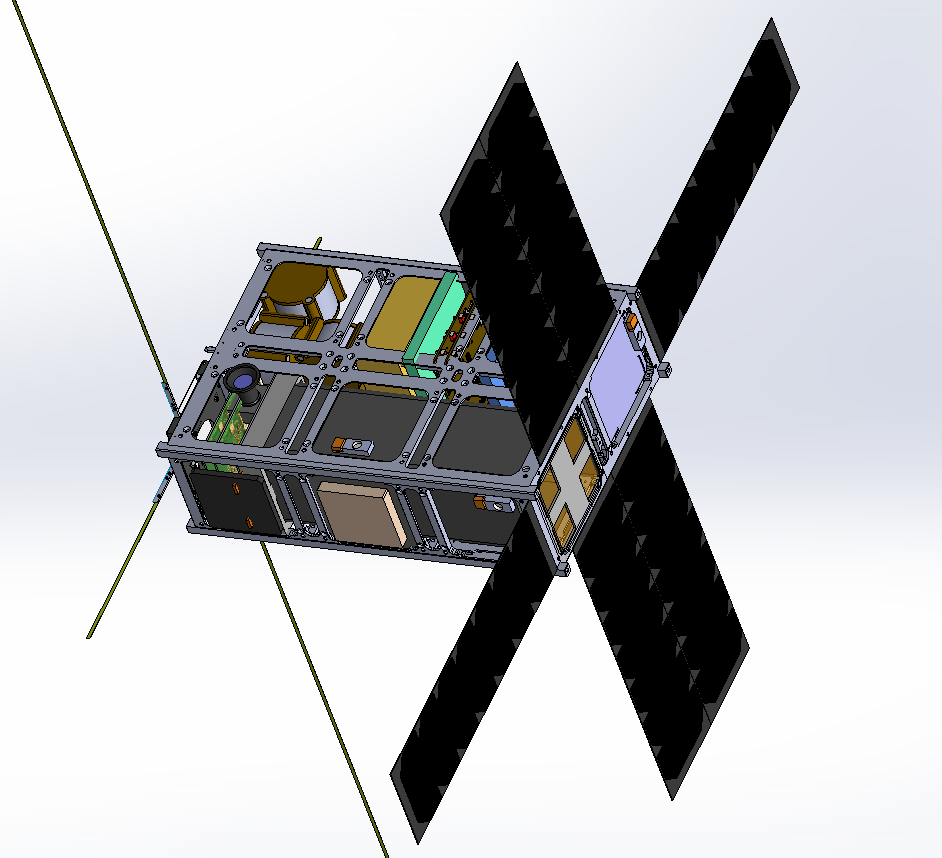}
   \caption{Preliminary satellite design.}
   \label{fig-sat}
 \end{figure}

 MiSolFA is a very compact detector, with mass around 1~kg, about
 3~dm$^3$ volume, and requiring less than 10~W.  Thus, it can be
 installed on a small micro-satellite.  There are few possible options
 for the platform, although not all can accomodate the full
 configuration.
 The preferred option is a 6-units satellite (Fig.~\ref{fig-sat})
 jointly developed by Clyde Space in Glasgow, Scotland and the Swiss
 Space Center (SSC) in Lausanne.  Alternatively, MiSolFA could be
 installed in the simplified configuration on a 3-units satellite.
 The 6-units platform can offer enough space for an X-ray imaging
 spectrometer, to be located inside a volume formed by 3 units in a
 row.  At the same time it is small enough to guarantee several
 possibilities for the launch, with an affordable cost.  This is
 essential to guarantee operation during the initial STIX data taking,
 at the next solar maximum.  At the same time, a 6-units satellite can
 also host two instruments, for example MiSolFA (without imaging)
 together with another detector.  Currently, the most likely candidate
 is an EUV radiometer developed by PMOD/WRC in Davos, Switzerland.
 % although an expression of interest was also received about a
 % low-energy polarimeter developed by IAPS in Rome, Italy.

 The main requirement from MiSolFA is a 3-axes stabilized satellite,
 pointing the Sun and possibly following a Sun-synchronous orbit.  The
 pointing stability should be within 1~deg, well within the capability
 of commercial attitude control systems.  Rather than attempting to
 improve the stability, the Aspect System is used to provide the
 precise position of the Sun in the field of view, which allows to
 correct for the motion of the optical axis during the offline image
 reconstruction.  Pointing to the Sun makes it easy to satisfy the
 power requirements by means of solar panels.  It also provides a
 quite stable situation from the thermal point of view, as one face is
 always the hot side and the opposite one is the cold side.  The best
 operating temperature for the Caliste units is $-20$~C, but the
 temperature stability is more important than the actual operating
 point.  Preliminary thermal studies indicate that it may be possible
 to stay within $\pm1$~C withouth active cooling if the temperature is
 not much below 0~C, although work is still in progress on this side.

 The onboard data processing shall guarantee the possibility of
 transmitting to ground entire flares, even in presence of very
 limited connectivity (9600 b/s for 20 min per day is typical of
 nanosatellites, and is used as a conservative estimate of the
 available bandwidth).  The simplest approach is to save the
 information about each individual photon (time, detector, pixel,
 energy) into a memory buffer.  Very large flares may require several
 hundreds MB, hence a 1 GB buffer shall be adopted.  As there are huge
 differences in photon flux between the thermal and non-thermal
 regions, it is foreseen to compress the data in a configurable way.
 Below a certain energy (which may depend on the flare intensity),
 photons will not be transmitted to ground individually.  Rather,
 histograms will be filled for each detector pixel, representing the
 photon counts in each energy bin for the considered (configurable)
 time interval.  At higher energies, single photons can be transmitted
 individually without impacting on the bandwidth requirements.
 Finally, to avoid saturation at the lowest energies, a thin layer of
 absorber (e.g.\ 0.5~mm Al) will be placed in front of each
 subcollimator.  In addition to provide a light-tight box enclosing
 the Caliste units, the absorber attenuates in a known way the flux of
 photons in the soft X-ray range, preventing the saturation of the
 read-out electronics.  No movable attenuator (like in RHESSI and
 STIX) is foreseen, as its motion would disturb the pointing of the
 small satellite.

 \section{Summary and conclusion}

 Despite from the great advances in our understanding of solar flares
 made possible by a dozen years observations by RHESSI, there are
 still quite a number of open questions.  The meachanisms of energy
 conversion from magnetic to kinetic are not fully understood, nor is
 the propagation of accelerated electrons in the solar corona.  In the
 next solar maximum period, STIX and MiSolFA will provide simultaneous
 X-ray observations of solar flares from two different points of view,
 using cross-calibrated photon detectors.  For the first time, it will
 be possible to measure the faint emission from the acceleration
 source, at the top of the loop in the corona, together with the much
 brighter footpoints, at the denser chromosphere.  This kind of
 observation, impossible with a single instrumend based on the moir\'e
 effect, is necessary to understand the electron acceleration and
 propagation, the heating of chromospheric plasma and its
 ``evaporation'' into the corona. In addition, the directivity of the
 X-ray emission as a function of the photon energy will be measured,
 allowing to test the hypothesis of beamed electron transport from the
 corona down to the chromosphere.  This kind of stereo observation
 will also be able to provide information on the source volume, which
 is needed when computing the energy budget.

\end{document}